\documentclass{article}
\usepackage{spconf,amsmath,graphicx}
\usepackage[T1]{fontenc}
\usepackage{multirow}
\usepackage{amssymb}
\usepackage{hyperref}
\usepackage{amsmath}
\usepackage[font=small,skip=4pt]{caption}
\DeclareMathOperator*{\argmax}{argmax}
\title{Reference-based Texture transfer for  Single  Image  Super-resolution of Magnetic Resonance images}
\name{\begin{tabular}{@{}c@{}}
Madhu Mithra K K$^{1\star}$ \thanks{$^{\star}$ Contributed equally} \qquad Sriprabha Ramanarayanan$^{1,2\star}$ \qquad Keerthi Ram$^{2}$  \\ Mohanasankar Sivaprakasam$^{1,2}$
\end{tabular}
}

\address{$^{1}$ Indian Institute of Technology Madras (IITM), India \\ $^{2}$ Healthcare Technology Innovation Centre (HTIC), IITM, India} 

\begin{document}
%\ninept
%
\maketitle
\begin{abstract}

Magnetic Resonance Imaging (MRI) is a valuable clinical diagnostic modality for spine pathologies with excellent characterization for infection, tumor, degenerations, fractures and herniations. However in surgery, image-guided spinal procedures continue to rely on CT and fluoroscopy, as MRI slice resolutions are typically insufficient.
Building upon state-of-the-art single image super-resolution, we propose a reference-based, unpaired multi-contrast texture-transfer strategy for deep learning based in-plane and across-plane MRI super-resolution.
We use the scattering transform to relate the texture features of image patches to unpaired reference image patches, and additionally a loss term for multi-contrast texture.
We apply our scheme in different super-resolution architectures, observing improvement in PSNR and SSIM for 4x super-resolution in most of the cases.

\end{abstract}
\section{Introduction}
\label{sec:intro}

Magnetic Resonance (MR) Imaging provides high value in spine clinical diagnosis, with standard classification and grading schemes for spinal pathologies built on MRI. Some of these rely on subtle findings specific to MRI such as end-plate abnormalities and high-intensity zones \cite{harada_imaging_2019}.
While spine MRI is unmatched for diagnosis and post-operative prognosis \cite{Kim2017}, surgery guidance is dominated by radiation imaging: CT and fluoroscopy. A primary requirement for intra-operative use is geometric accuracy, which is typically lower in MRI due to acquisition of thicker slices.
Super-resolution (SR) could be an enabler for use of MRI intra-operatively in spine procedures.
We take up deep learning super-resolution as a direct means of improving spine MRI resolution, leveraging the unique possibility in MRI of acquiring multiple contrast images and with different slicing direction. 

\textbf{Related work:} Some state-of-the-art single image super-resolution (SISR) techniques for natural images include Residual Channel Attention network (RCAN) \cite{Zhang2018} and Residual Feature Distillation Network (RFDN) \cite{Liu2020}. Recent work in MRI super-resolution include \cite{jurek_cnn-based_2020}, which evaluated SSIM improvement in SRCNN for brain MRI, and showed superior reconstruction compared to fusion of thick slices of different anatomical orientation. Multi-contrast super-resolution has been explored in \cite{lyu_multi-contrast_2020} where synergizing multiple contrasts was shown to achieve better super-resolution results, using a composition of loss terms including MSE, adversarial, perceptual and textural loss. 
Other approaches in literature include residual deep learning \cite{He2020} and use of cross-plane self-similarity in MRI \cite{du_super-resolution_2020}.

Peculiar to MRI, the different MRI contrast images have similar texture patterns. State-of-art SISR networks like RCAN, RFDN, can be augmented to utilize this. Reference based SISR methods have been effective in leveraging  texture details from reference images to assist resolution enhancement \cite{Zhang2019}. But VGG \cite{vgg2014} type feature-maps which are used in natural image SR would not be suited, as they are known to be sensitive to lower order image characteristics, capturing structure more than texture. The learnt filters are also not guaranteed to be orthogonal, making comparison and matching using simplistic distance metrics questionable. 
\begin{figure}[htb]
  \centering
  \centerline{\includegraphics[width=1\linewidth]{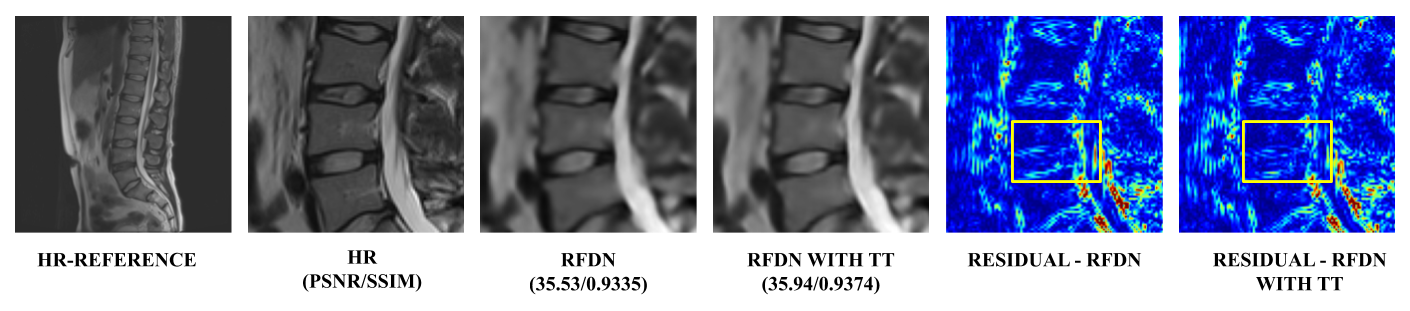}}
\caption{Proposed SISR with texture transfer}
\label{fig:front}
\end{figure}
In our work we emphasize on using textures of multi-contrast MRI images as reference for super-resolution.
Texture transfer (TT) attempts to replace the texture features of Low Resolution (LR) MRI input with matched textures of the High Resolution (HR) reference. Before comparing the LR textures with the reference textures, the reference images are degraded by blurring so that matching operates in similar sampling frequency.

In place of VGG feature maps like in prior works, we use wavelet scattering to extract multiscale texture features and use a texture-based loss function for texture transfer.
Our approach achieves texture transfer by extracting wavelet scattering features, compared using dense matching to an unpaired HR reference, and combining texture information from the matched reference. Fig. \ref{fig:front} shows a quick overview of the proposed reference-based texture transfer approach.

We summarize our contributions as follows:
\begin{enumerate}
\item A module applicable onto baseline architectures like SRCNN \cite{Dong2016}, RCAN, RFDN, providing a reference-based, unpaired, texture transfer strategy using wavelet scattering and texture loss, utilizing multi-contrast MRI.
\item We explore the effect of using different kinds of reference images in improving the resolution quality; for instance, use of axial reference images to train model for sagittal LR-HR.
% \item We compare 3 existing SISR architectures: SRCNN, RCAN, RFDN, axial and sagittal, with different references, Results show improvement for most cases across models improving through-plane resolution (4x) both quantitatively and qualitatively.
\item We compare 3 existing SISR architectures: SRCNN, RCAN, RFDN on sagittal and axial MRI input LR images with different types of MRI reference images.  Results show improvement for most cases across models improving the in-plane and through-plane resolution (4x) both quantitatively and qualitatively.
\end{enumerate} 

\begin{figure*}[htb]

\begin{minipage}[b]{0.68\linewidth}
  \centering
  \centerline{\frame{\includegraphics[width=.9\linewidth]{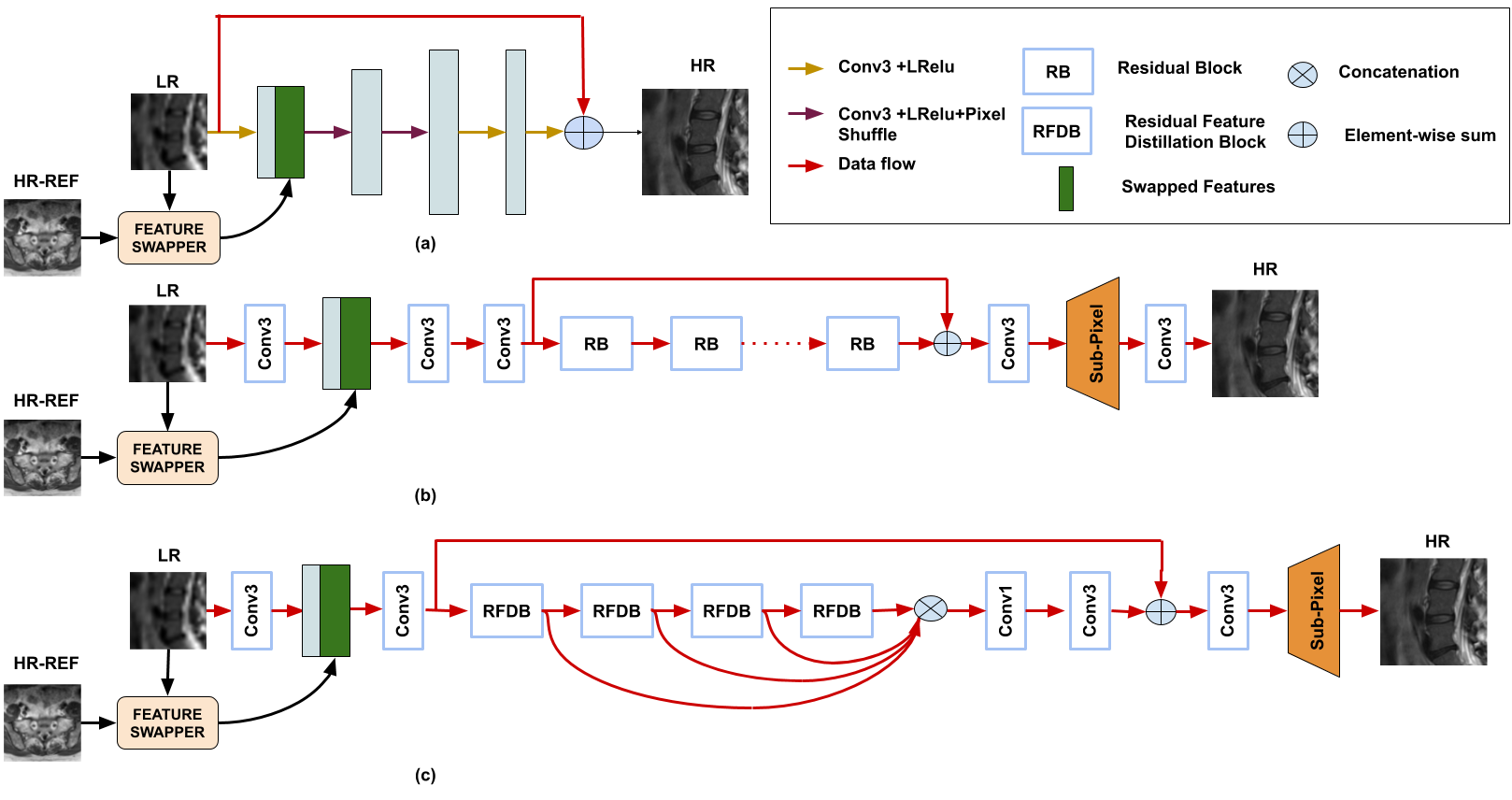}}}
%  \vspace{2.0cm}
  \centerline{ (i)  (a)\, SRCNN, (b)\,RCAN and (c)\,RFDN with reference-based texture transfer}\medskip
\end{minipage}
\begin{minipage}[b]{0.28\linewidth}
  \centering
  \centerline{\frame{\includegraphics[width=1.2\linewidth,height=1.15\linewidth]{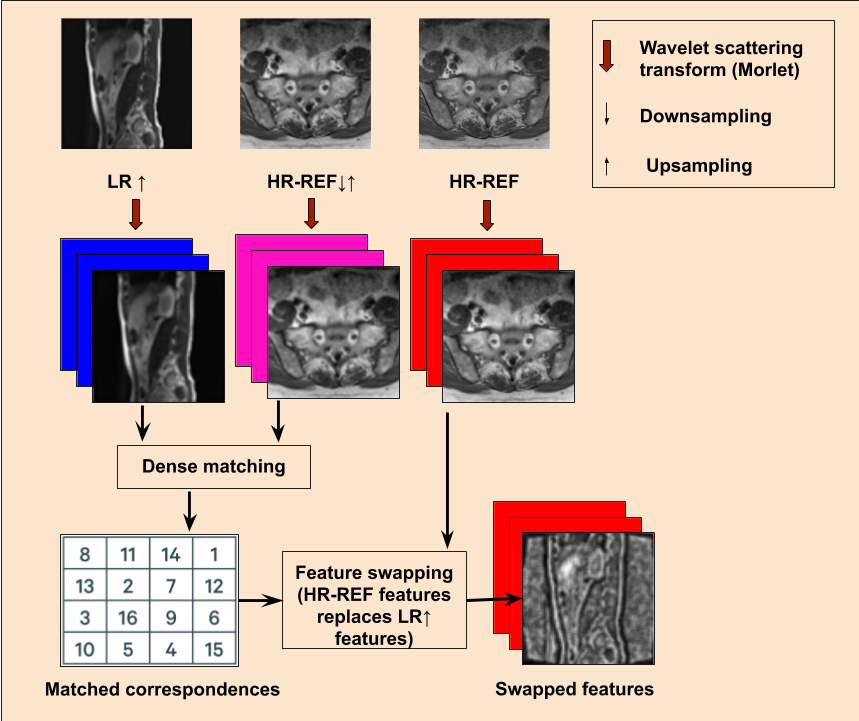}}}
%  \vspace{2.0cm}
 \centerline{(ii) Feature swapper}\medskip
  \centering
\end{minipage}
\caption{SISR Architectures of with reference-based texture transfer}
\label{fig:flowchart}
\end{figure*}

\section{SUPER RESOLUTION THROUGH TEXTURE TRANSFER}
\label{sec:Superresolution}

\subsection{Background study for texture transfer}
\label{ssec:Method}
The concept of reference-based texture transfer has been inspired from the idea of neural style transfer using VGG features \cite{Zhang2019} in SISR. For MR images, texture details can be retrieved from the images acquired across multiple protocols and hence can be used for texture transfer. Wavelet is one of the best ways to decompose the image into various frequency bands \cite{Livens1997}, which can be effectively used in texture extraction. Hence, in our approach of reference-based texture transfer, we compute the wavelet transform of the LR image to be super resolved and the high resolution reference to extract texture features. In the transformed domain, feature swapping is performed wherein the texture features of LR is replaced with the best matched texture features of the reference by dense comparison. 
To minimize the difference between the swapped texture features and the texture features of super resolved output, we include a texture loss term in the loss function.
% To minimize the difference between the texture features of the swapped reference and the super resolved output, we include a texture loss term in the loss function.

In the proposed method, Morlet wavelet transform \cite{Andreux2020} is used for computing the texture features of the image. Morlet wavelet filters are very similar to Gabor filters \cite{Livens1997}, which have been widely used in texture extraction. Wavelet coefficients are computed at different scales, where scale denotes the level of decomposition of the image. 

Additionally, in the proposed approach based on wavelet transform, wavelet filters at different rotation angles are considered to ensure rotation invariance in the computed features.

 \subsection{Feature Swapping}
\label{ssec:TT}
 There are two main steps in feature swapping - Dense matching and texture transfer. 
The LR image ($I_{LR}$) is first upsampled to the actual size using bicubic interpolation.
The HR reference images are down-sampled and up-sampled using bicubic interpolation to obtain blurred reference images so as to bring them to the same frequency bandwidth as the LR input. The process of feature swapping is illustrated in Fig. \ref{fig:flowchart}. The wavelet scattering coefficients are obtained for upsampled LR input ($I_{LR}\uparrow$), HR reference ($I_{Ref}$) and blurred HR reference ($I_{Ref}\downarrow\uparrow$). The features of LR input and that of blurred reference are sampled into small dense patches of size 3x3 with a stride of 1. Each texture patch of $I_{LR}\uparrow$ is compared with every other texture patches of $I_{Ref}\downarrow\uparrow$, to get the corresponding reference texture patch index that gives the highest correlation. 
To represent this mathematically, let the sampling function be given as $P$ and the similarity map between the $i^{th}$ input texture patch and the $j^{th}$ reference texture patch can be given as $s(i,j)$. The similarity between two patches is calculated as a dot product.
\begin{equation}
    s(i,j) = \langle\, P_{i}(\phi_{J,L}(I_{LR}\uparrow))\, , \,  P_{j}(\phi_{J,L}(I_{Ref}\downarrow\uparrow))\, \rangle
\end{equation}
Here $\phi_{J,L}$ denotes the texture feature extractor with scale $J$ and number of rotation angles $L$.
The index of the reference texture patch that gives the maximum similarity is chosen as the texture for that particular LR image texture patch. 
\begin{equation}
    S(i) = \argmax_j \,s(i,j)
\end{equation}
$S(i)$ is the index of reference texture patch of maximum correlation. This operation (called as dense matching operation) is repeated for every LR input patch to obtain the best matching correspondences in texture and can be formulated as,
% for the complete input texture features

%We call this operation, dense matching which is formulated as follows.
\begin{equation}
    M(J,L) = D( \phi_{J,L}(I_{LR}\uparrow) , \phi_{J,L}(I_{Ref}\downarrow\uparrow)
\end{equation}
Here, $D$ denotes the dense matching operation and $M(J,L)$ is the matching correspondences between the features of LR input and that of blurred reference.  

Based on the correspondences, the patches of texture features of HR reference are placed at the appropriate positions. We call this operation of transfer of texture information from the HR reference to the LR input as texture transfer.
%, carried out with the obtained matching correspondences can be formulated as below.
\begin{equation}
    F = H( \phi_{J,L}(I_{Ref}), M(J,L))
\end{equation}
where, $H$ denotes the texture transfer operation and $F$ denotes the transferred features that can be concatenated to the SISR model during training at a specified layer (typically at one of the initial layers) of the model based on its architecture.

\subsection{Loss functions}
\label{ssec:loss}
We used three loss functions in the proposed approach namely the reconstruction loss in the image domain, perceptual loss and texture loss in the feature domain. 
% For the reconstruction loss, $L_1$ is preferred over Mean Squared Error (MSE) as the former gives a sharper image \cite{Zhao2017}. 
The reconstruction loss $L_{rec}$ is given as, 
\begin{equation}
    L_{rec} = \|I_{SR} - I_{HR}\|_1
\end{equation}
where $I_{SR}$ is the super resolved (or predicted) image and the $I_{HR}$ is the HR ground-truth image.
The perceptual loss component is applied \cite{Johnson2016} to improve the visual quality of the image by enforcing similarity with respect to the ground truth in the feature space. Typically, we adopt the ReLU5-1 layer of VGG19 \cite{Liu2015}. It can be given by, 
\begin{equation}
    L_p = \frac{1}{S} \sum_{i=1}^{N}\| \theta_i (I_{SR}) - \theta_i (I_{HR}) \|_F
\end{equation}
where $S$ is the features size (width * height * number of channels), $N$ is the number of channels and $\theta_i$ denotes the $i^{th}$ channel of VGG19. 
The third loss function enforces texture similarity between the swapped features of LR (from reference images) and the super resolved image. The texture loss is computed by taking Frobenius norm of Gram matrix of the extracted texture features.
\begin{equation}
    L_t = \frac{1}{W} \| Gr(\phi_{J,L} (I_{SR})) - Gr(F) \|_{F}
\end{equation}
Here, $F$ and $\phi_{J,L}$ are defined in Section \ref{ssec:TT}, $W$ is the feature size and $Gr(.)$ calculates the Gram matrix, which helps in computing the texture information \cite{Gatys2016}.

\section{EXPERIMENTS}
\label{sec:Experiments}
\subsection{Data preparation}
\label{ssec:Dataset}
We have used two publicly available datasets for training and validation - Zenodo\cite{zenodo} and Dataset1
from SpineWeb \cite{noauthor_spineweb_nodate}. 
We have chosen 23 volumes of T2-weighted turbo spin echo 3D MR of Zenodo and 8 MRI T1-weighted volumes of SpineWeb Dataset1 for training. The HR patches are created by randomly cropping twenty 64 x 64 patches from each MRI 2D slice making a total of 14000 patches (10920 sagital view patches from Zenodo and 3080 sagittal SpineWeb patches together) for training. For testing, 312 sagittal view images of Zenodo and 316 axial view images of SpineWeb are used. Wilcoxon signed-rank test with an alpha of 0.05 is used to assess statistical significance.
\renewcommand{\arraystretch}{0.9}
\begin{table*}[htb]
    \small
    \centering
    \begin{tabular}{|c|c|c|c|c|}
    \hline
    \multirow{2}{4em}{Model name} & \multicolumn{2}{c|}{Sagittal view data} & \multicolumn{2}{c|}{Axial view data} \\\cline{2-5}
    
    & PSNR (dB) & SSIM & PSNR (dB) & SSIM \\
    \hline
 %    Bilinear & 34.4679 & 0.9126 & 28.2666 & 0.8691  \\
  %  \hline
    Bicubic & 35.1682 & 0.9221 & 28.9117 & 0.8827  \\
    \hline
     SRCNN ($L_1+L_p$) & 37.4923 & 0.9483 & 29.7490 & 0.8985 \\
     \hline
     SRCNN ($L_1+L_p+L_t$)-Sag & \textbf{38.0789} & \textbf{0.9550} & \textbf{30.0975} & \textbf{0.9041}   \\ 
     \hline
     SRCNN($L_1+L_p+L_t$)-Ax & \textbf{38.1703} & \textbf{0.9557} & \textbf{30.1647} & \textbf{0.9050} \\
     \hline
      RCAN ($L_1+L_p$) & 38.4679 & 0.9596 & 30.3360 & 0.9066 \\
     \hline
     RCAN ($L_1+L_p+L_t$)-Sag &  \textbf{38.5239} & \textbf{0.9601} & \textbf{30.3383} & \textbf{0.9072} \\
     \hline
     RCAN ($L_1+L_p+L_t$)-Ax & \textbf{38.5320} & \textbf{0.9603} & \textbf{30.3883} & \textbf{0.9078} \\
     \hline
     RFDN ($L_1+L_p$) & 38.4368 & 0.9586 & 30.3409 & 0.9073 \\
     \hline
     RFDN ($L_1+L_p+L_t$)-Sag &  \textbf{38.5021} & \textbf{0.9587} & 30.1427 & 0.9055 \\
     \hline
     RFDN ($L_1+L_p+L_t$)-Ax & \textbf{38.5497} & \textbf{0.9594} & 30.2344 & 0.9066 \\
     \hline
    \end{tabular}
    \caption{Quantitative comparison of SRCNN, RCAN and RFDN without and with texture transfer (metrics improvements shown in bold). Sagittal view data - Zenodo images, Axial view data - SpineWeb images. Sagittal and axial texture transfer models end with '-Sag' and '-Ax'}
    \label{tab:1}
\end{table*}
\begin{figure*}[htb]

  \centering
  \centerline{\includegraphics[width=0.95\linewidth]{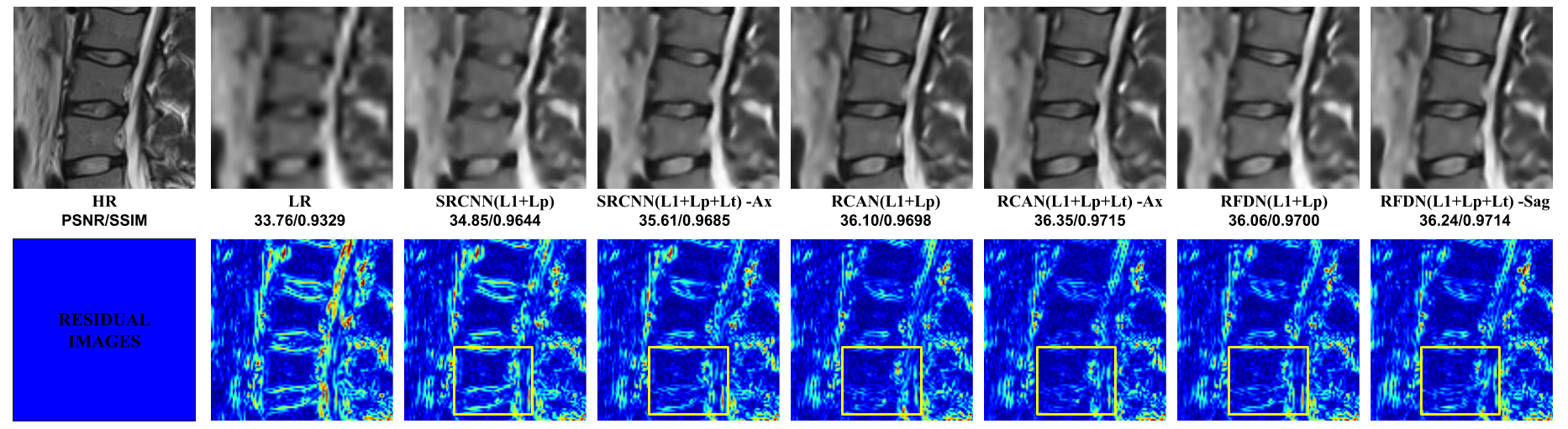}}
\caption{Visual comparison of methods SRCNN, RCAN and RFDN with and without texture transfer}
\label{fig:res}
\end{figure*}
\subsection{Training details with and without texture transfer}
\label{ssec:training}
The models chosen for training are the pre-existing SISR models -  SRCNN, RCAN and RFDN with and without texture transfer. 
% For training, the patches obtained from Zenodo dataset and Spineweb dataset were combined and shuffled making it 14000 patches. 
For each model we conduct three trainings - one without TT with only $L_1$ and $L_p$ loss components.
% only reconstruction and perceptual loss components
(denoted by model name followed by ($L_1+L_p$)) and the other two with TT based on two different sets of reference images (sagital and axial reference, denoted by model name followed by ($L_1+L_p+L_t$) with suffix '-Sag' or '-Ax' respectively ). 
Before training the baseline models with TT, feature swapping followed by concatenation of swapped features to one of the initial layers of the model is performed. 
The two types of reference images used in our study are - 1. Sagital TT: Four sagittal reference images from Zenodo for sagittal reference based training 2. Axial TT: Five Axial reference images for axial reference based training.
For texture extraction we use the Morlet wavelet transform with J = 2 and L = 8.
We have used Adam optimizer with learning rate 1e-4 for all the models. The batch size is chosen to be 9 for SRCNN and RFDN, and 16 for RCAN. 
All the implementations are carried out using Pytorch.

\section{RESULTS}
\label{sec:typestyle}
The three convolutional neural network (CNN) models (SRCNN, RCAN and RFDN) are chosen based on the complexity of the architectures. The SRCNN is the simplest of the three models. RCAN is one of the state-of-the-art deep models with residual blocks and channel attention layers. RFDN is one of the recent models having multiple residual feature distillation blocks.
% The three CNN models are trained with reconstruction loss and perceptual loss, which are taken as our baselines such that the effect of texture transfer can be seen distinctly.  
Table \ref{tab:1} shows the quantitative comparison of the proposed approach. From the table, the following observations are noted. 1) The performance improvement of the proposed texture transfer has been consistent across the SISR baseline CNN models in most of the cases under study (highlighted in bold). 2) The axial texture transfer results are better than sagittal texture transfer. This is an important observation which emphasizes the purpose of reference based texture transfer. The reason could be that the texture features taken from high resolution reference image (in this case axial reference images which are T1-weighted images of voxel size 0.3 mm x 0.3 mm) are richer as compared to those with lower resolution. 3) The performance improvements are consistent across models for sagittal texture transfer. The resulting metrics are found to be statistically significant (pvalue $<$ 0.05). For axial texture transfer, the SRCNN and RCAN show improvements while RFDN has shown equivalent performance in SSIM and a reduction in performance in PSNR. Since RFDN has multiple dense connections with multi-distillation blocks, we believe that the layer through which TT is done could be a reason in influencing the overall performance. As a future work, we are studying more CNN-based SISR models to understand the consistency of different texture transfer strategies across layers of the models. 
%Also, the idea can be applied on the phantom studies to verify its  correctness.
Fig. \ref{fig:res} shows the inter-vertebral sections of the lumbar spine anatomy. The qualitative comparison of bicubic interpolation method and models with and without TT is shown.  From figure it is clear that the linear structures (highlighted with a box) in the discs are reconstructed with better agreement to the groundtruth as compared with the baseline CNN models. 
\section{CONCLUSION}
\label{sec:typestyle}
Towards enabling the use of MRI in spine surgery intra-operative guidance, we take up single image super-resolution from clinical MRI, and demonstrated our novel method of reference-based texture transfer, implemented in 3 baseline architectures, and evaluated on two public datasets for multi-contrast SR in different slicing directions. Comparing with the baseline, the proposed texture transfer with texture loss exhibits good improvements in the performance evaluation metrics together with visual quality improvement. The observations show that the proposed neural texture transfer can be a potential source of improving the resolution of MRI images. 
\section{Compliance with Ethical Standards}
\label{sec:cwe}
This research study was conducted retrospectively using human subject data made available in open access by Zenodo \cite{zenodo} and SpineWeb - Dataset1 \cite{noauthor_spineweb_nodate}. Ethical approval was not required as confirmed by the license attached with the open access data.
\section{Acknowledgements}
\label{sec:ack}
No funding was received for conducting this study. The authors have no relevant financial or non-financial interests to disclose.

\bibliographystyle{IEEEbib}
\bibliography{bib1}

\begin{thebibliography}{10}

\bibitem{harada_imaging_2019}
Garrett~K Harada, Zakariah~K Siyaji, Sadaf Younis, Philip~K Louie, Dino
  Samartzis, and Howard~S An,
\newblock ``Imaging in {Spine} {Surgery}: {Current} {Concepts} and {Future}
  {Directions},''
\newblock {\em Spine Surgery and Related Research}, vol. 4, no. 2, pp. 99--110,
  Nov. 2019.

\bibitem{Kim2017}
Seon~Jeong Kim, Sang~Hoon Lee, and et~al. Chung,
\newblock ``{Magnetic resonance imaging patterns of post-operative spinal
  infection: Relationship between the clinical onset of infection and the
  infection site},''
\newblock {\em Journal of Korean Neurosurgical Society}, vol. 60, no. 4, pp.
  448--455, 2017.

\bibitem{Zhang2018}
Yulun Zhang, Kunpeng Li, Kai Li, Lichen Wang, Bineng Zhong, and Yun Fu,
\newblock ``Image super-resolution using very deep residual channel attention
  networks,''
\newblock in {\em Proc. IEEE {ECCV}}. 2018.

\bibitem{Liu2020}
Jie Liu, Jie Tang, and Gangshan Wu,
\newblock ``Residual feature distillation network for lightweight image
  super-resolution,''
\newblock {\em arXiv:2009.11551 [cs, eess]}, 2020.

\bibitem{jurek_cnn-based_2020}
Jakub Jurek, Marek KociÅ„ski, Andrzej Materka, Marcin Elgalal, and Agata
  Majos,
\newblock ``{CNN}-based superresolution reconstruction of {3D} {MR} images
  using thick-slice scans,''
\newblock {\em Biocybernetics and Biomedical Engineering}, vol. 40, no. 1, pp.
  111--125, Jan. 2020.

\bibitem{lyu_multi-contrast_2020}
Q.~Lyu, H.~Shan, C.~Steber, C.~Helis, C.~Whitlow, M.~Chan, and G.~Wang,
\newblock ``Multi-{Contrast} {Super}-{Resolution} {MRI} {Through} a
  {Progressive} {Network},''
\newblock {\em IEEE Transactions on Medical Imaging}, vol. 39, no. 9, pp.
  2738--2749, Sept. 2020.

\bibitem{He2020}
Xiuxiu He, Yang Lei, Yabo Fu, Hui Mao, Walter~J. Curran, Tian Liu, and Xiaofeng
  Yang,
\newblock ``Super-resolution magnetic resonance imaging reconstruction using
  deep attention networks,''
\newblock in {\em SPIE Medical Imaging 2020: Image Processing}. Mar. 2020,
  {SPIE}.

\bibitem{du_super-resolution_2020}
Jinglong Du, Zhongshi He, Lulu Wang, Ali Gholipour, Zexun Zhou, Dingding Chen,
  and Yuanyuan Jia,
\newblock ``Super-resolution reconstruction of single anisotropic 3d {MR}
  images using residual convolutional neural network,''
\newblock {\em Neurocomputing}, vol. 392, pp. 209--220, June 2020.

\bibitem{Zhang2019}
Zhifei Zhang, Zhaowen Wang, Zhe Lin, and Hairong Qi,
\newblock ``{Image super-resolution by neural texture transfer},''
\newblock {\em Proc. IEEE CVPR}, 2019.

\bibitem{vgg2014}
Karen Simonyan and Andrew Zisserman,
\newblock ``Very deep convolutional networks for large-scale image
  recognition,''
\newblock {\em arXiv 1409.1556}, 09 2014.

\bibitem{Dong2016}
Chao Dong, Chen~Change Loy, Kaiming He, and Xiaoou Tang,
\newblock ``Image super-resolution using deep convolutional networks,''
\newblock {\em {IEEE} Transactions on Pattern Analysis and Machine
  Intelligence}, vol. 38, no. 2, pp. 295--307, Feb. 2016.

\bibitem{Livens1997}
S.~Livens, P.~Scheunders, G.~{Van de Wouwer}, and D.~{Van Dyck},
\newblock ``{Wavelets for texture analysis, an overview},''
\newblock {\em IEE Conference Publication}, , no. 443 pt 2, pp. 581--585, 1997.

\bibitem{Andreux2020}
Mathieu Andreux, Tom{\'{a}}s Angles, Georgios Exarchakis, and et~al.
  Leonarduzzi,
\newblock ``{Kymatio: Scattering transforms in python},''
\newblock {\em Journal of Machine Learning Research}, vol. 21, no. 2012, pp.
  2012--2017, 2020.

\bibitem{Johnson2016}
Justin Johnson, Alexandre Alahi, and Li~Fei-Fei,
\newblock ``Perceptual losses for real-time style transfer and
  super-resolution,''
\newblock in {\em Proc. IEEE {ECCV}}. 2016.

\bibitem{Liu2015}
Shuying Liu and Weihong Deng,
\newblock ``Very deep convolutional neural network based image classification
  using small training sample size,''
\newblock in {\em 2015 3rd {IAPR} Asian Conference on Pattern Recognition
  ({ACPR})}. Nov. 2015, {IEEE}.

\bibitem{Gatys2016}
Leon~A. Gatys, Alexander~S. Ecker, and Matthias Bethge,
\newblock ``{Image Style Transfer Using Convolutional Neural Networks},''
\newblock in {\em Proc. IEEE CVPR}, 2016.

\bibitem{zenodo}
``Zenodo: [http://dx.doi.org/10.5281/zenodo.22304],'' .

\bibitem{noauthor_spineweb_nodate}
``{SpineWeb} : {Main} / {Datasets} : [http://spineweb.digitalimaginggroup.ca/
  index.php ?n=main.datasets],'' .

\end{thebibliography}

\end{document}